\documentclass[sigconf,authorversion]{acmart}

\AtBeginDocument{%
  \providecommand\BibTeX{{%
    \normalfont B\kern-0.5em{\scshape i\kern-0.25em b}\kern-0.8em\TeX}}}

\copyrightyear{2024} 
\acmYear{2024} 
\setcopyright{acmlicensed}\acmConference[CUI '24]{ACM Conversational User Interfaces 2024}{July 8--10, 2024}{Luxembourg, Luxembourg}
\acmBooktitle{ACM Conversational User Interfaces 2024 (CUI '24), July 8--10, 2024, Luxembourg, Luxembourg}
\acmDOI{10.1145/3640794.3665539}
\acmISBN{979-8-4007-0511-3/24/07}

\usepackage{graphicx}
\usepackage{url}
\usepackage{multirow}
\usepackage{color}
\usepackage{enumitem}
\usepackage{subcaption}
\usepackage{algorithm}

\usepackage{algorithmic}
\newcommand\algorithmicprocedure{\textbf{procedure}}
\newcommand{\algorithmicendprocedure}{\algorithmicend\ \algorithmicprocedure}
\makeatletter
\newcommand\Procedure[3][default]{%
  \ALC@it
  \algorithmicprocedure\ \textsc{#2}(#3)%
  \ALC@com{#1}%
  \begin{ALC@prc}%
}
\newcommand\EndProcedure{%
  \end{ALC@prc}%
  \ifthenelse{\boolean{ALC@noend}}{}{%
    \ALC@it\algorithmicendprocedure
  }%
}
\newenvironment{ALC@prc}{\begin{ALC@g}}{\end{ALC@g}}
\makeatother

\newcommand{\citeanon}[1]{
\ifdefined\anonymous
	[anonymous]%
\else
	\citep{#1}%
\fi
}

\makeatletter
\newcommand*{\rom}[1]{\expandafter\@slowromancap\romannumeral #1@}
\makeatother

\begin{document}
\fancyhead{}

\title{Identifying Breakdowns in Conversational Recommender Systems using User Simulation} 

\author{Nolwenn Bernard}
\affiliation{%
  \institution{University of Stavanger}
  \city{Stavanger}
  \country{Norway}
}
\email{nolwenn.m.bernard@uis.no}

\author{Krisztian Balog}
\affiliation{%
  \institution{University of Stavanger}
  \city{Stavanger}
  \country{Norway}
}
\email{krisztian.balog@uis.no}

\begin{abstract}
We present a methodology to systematically test conversational recommender systems with regards to conversational breakdowns. It involves examining conversations generated between the system and simulated users for a set of pre-defined breakdown types, extracting responsible conversational paths, and characterizing them in terms of the underlying dialogue intents. 
User simulation offers the advantages of simplicity, cost-effectiveness, and time efficiency for obtaining conversations where potential breakdowns can be identified.
The proposed methodology can be used as diagnostic tool as well as a development tool to improve conversational recommendation systems.
We apply our methodology in a case study with an existing conversational recommender system and user simulator, demonstrating that with just a few iterations, we can make the system more robust to conversational breakdowns.

\end{abstract}



\maketitle

\section{Introduction}

Conversational recommender systems (CRSs) aim to provide personalized recommendations to user through multi-turn conversations~\citep{Jannach:2021:CSUR,Gao:2021:AIOpen}.
However, ensuring the robustness and effectiveness of these systems under any and all of the possible situations encountered while engaging with users remains a critical challenge~\citep{Brabra:2022:TCDS}. 
We define a \emph{breakdown} as a moment in the conversation where the flow discontinues or even stops. This includes system failures and unexpected/irrelevant replies from the CRS.
Examples include the CRS recommending an item without knowing what the user wants or prefers (i.e., recommendation before elicitation of the need and preferences) and the CRS repeating itself over and over.
Previously, the dialogue breakdown detection challenge~\citep{Higashinaka:2016:LREC} motivated research on breakdown detection in chat-oriented dialogues. During the challenge, an annotated dialogue corpora was provided to the participants. Note that the set of labels used does not give information on the specific type of breakdown, despite the availability of at least one detailed breakdown taxonomy~\citep{Higashinaka:2015:SIGDIAL}. Consequently, most of the proposed solutions were supervised classification approaches determining if a (potential) breakdown occurred or not.
Our work focuses on recommendation dialogues, i.e., task-oriented dialogues, that are not annotated. Furthermore, we aim to identify specific breakdowns in dialogues. Therefore, we propose a novel methodology to identify breakdowns and assess the robustness of an existing CRS via user simulation.

Leveraging user simulation has several advantages.
First, it is a simple, cost-effective, and efficient solution to test a CRS that supplements human evaluation~\citep{Zhang:2020:KDD}.
Second, it allows for a comprehensive assessment of a CRS's abilities in various scenarios and even allows for the simulation of the behavior of users with different characteristics (e.g., impatient, selective).
In particular, our methodology identifies conversational paths (i.e., sequences of intents) that lead to conversational breakdowns. To achieve this, a set of conversations between the CRS and simulated users is analyzed with regards to a set of pre-defined breakdowns.
As a starting point, we present detectors for three specific breakdowns to demonstrate our methodology: \emph{system failure} (bugs), \emph{dialogue of the deaf} (communication breakdowns between parties), and \emph{conversational flow discontinuation} (disruptions with regards to a predefined interaction model).
The identification of breakdowns in the conversational flow can provide insights to improve the robustness and effectiveness of the CRS. 
Furthermore, if changes to the behavior of the CRS can be made (e.g., either by updating the source code or providing it with further training examples), then the process can be repeated iteratively.

To demonstrate the usefulness of our approach, we present a case study with an existing CRS and user simulator, in which the conversational breakdown detection is performed on four subsequent versions of the CRS. More specifically, after each breakdown detection, a modification is applied to the CRS based on insights extracted (i.e., conversational patterns) with the goal to reduce the number of breakdowns, hence, improve the CRS.
The results show that modifying the CRS with regards to one type of breakdown reduces its presence, while also affecting the number of other types of breakdowns detected. 
Moreover, we note that some breakdowns stem from imperfections in the user simulator. We demonstrate that our methodology can help improve the user simulator, parallel to the CRS, thereby making it a more robust and effective tool.

In summary, the main contributions of this work are twofold.
First, we propose a methodology to identify conversational breakdowns in CRSs.
Second, we present a case study where we apply this approach with an existing open-sourced CRS and user simulator. More specifically, we show how we improve the CRS based on the breakdowns identified.

\section{Related Work}
\label{sec:related}

Conversational recommender systems can be built following task-oriented dialogue system architectures or as end-to-end trainable systems~\citep{Gao:2021:AIOpen}.
In the former case, the \emph{dialogue policy} is a central component, which determines the system's response and action to a user utterance.
Hence, a well-designed dialogue policy allows the CRS to seamlessly handle different situation such as unexpected actions from users and misunderstandings. 
The dialogue policy may be rule-based~\citep{Ardissono:2004:CI,Warnestaal:2005:workshop} or model-based, commonly trained using reinforcement learning~\citep{Scheffler:2001:NAACL,Mahmood:2009:HT,Sun:2018:SIGIR,Lei:2020:WSDM}.
Our methodology remains independent of the specific architecture of the CRS, and can be used with any of the aforementioned solutions.

\begin{table*}
    \centering
    \caption{Categorization of breakdowns identified from the references~\citep{Higashinaka:2015:SIGDIAL,Higashinaka:2021:SIGDIAL,Moller:2007:INTERSPEECH,Oulasvirta:2006:PQS}. The categories considered in this work are marked with $^\dagger$.}
    \label{tab:breakdowns_categories}
    \begin{tabular}{p{0.1\textwidth}|p{0.42\textwidth}|p{0.42\textwidth}}
        \hline
        \textbf{Category} & \textbf{Description} & \textbf{Breakdowns} \\ \hline 
        Linguistic & Breakdowns occurring due to language issues in an utterance & Syntactic error, semantic error, uninterpretable, wrong information, and command-level error \\ \hline 
        Contextual$^\dagger$ & Breakdowns occurring when the utterance is not appropriate with regards to the context & Excess/lack of information, non-understanding, no-relevance, unclear intention, misunderstanding, excess/lack of proposition, contradiction, self-contradiction, non-relevant topic, topic switch error, and repetition \\ \hline 
        Social & Breakdowns related to social norms & Lack of common ground, lack of common sense, and lack of sociality \\ \hline 
        Functional$^\dagger$ & Breakdowns related to the absence of a function or the inability to perform a function & Goal-level error, recognition-level error, and concept-level error \\ \hline 
    \end{tabular}
\end{table*}

Breakdowns in conversations have been studied in the field of dialogue systems and conversational analysis. It led to the development of different taxonomies of breakdowns~\citep{Higashinaka:2021:SIGDIAL,Higashinaka:2015:SIGDIAL,Moller:2007:INTERSPEECH,Oulasvirta:2006:PQS} and associated repair strategies~\citep{Benner:2021:ICIS}, i.e., strategies used to overcome a breakdown.
We group the breakdowns identified in~\citep{Higashinaka:2015:SIGDIAL,Higashinaka:2021:SIGDIAL,Moller:2007:INTERSPEECH,Oulasvirta:2006:PQS} into four categories which represent different aspects of a conversation (Table~\ref{tab:breakdowns_categories}).
In addition to these taxonomies, the BETOLD~\citep{Terragni:2022:CAI} and Dialogue Breakdown Detection Challenge (DBDC)~\citep{Higashinaka:2016:LREC} datasets were released to facilitate the development of breakdown detection methods. It is worth pointing out that these datasets are annotated with a small set of labels, which does not cover all types of breakdown mentioned in previous taxonomies. For example, in DBDC, there are three labels: not a breakdown, possible breakdown, and breakdown, which are given to each system utterance.
While, in BETOLD, the labels are binary and at the conversation level. The dialogue breakdown detection challenge~\citep{Higashinaka:2016:LREC} ran for several years (between 2016 and 2019) and resulted in the development of different approaches to tackle this tasks in the context of chat-oriented dialogues most of them being based on supervised learning~\citep{Hori:2019:CSL,Higashinaka:2016:LREC}, e.g., training of a recurrent neural network or support vector machine.
Instead, in this paper, we focus on task-oriented dialogues.
Although, chat-oriented and task-oriented dialogues have different characteristics, one can wonder if breakdown detection methods are generalizable. \citet{Lopes:2017:DSTC} studied this question by comparing the results of breakdown detection in task- and chat-oriented dialogues using the same set of features, e.g., cosine similarity between turns and bag-of-words representation of utterances. Based on this comparison, the author concludes that the hypothesis of a generic set of features to detect breakdowns ``is true to a certain extent'' due to the difference in interpretation of the features.
In this work, we take a step forward and propose a methodology to detect specific types of breakdown instead of merely identifying the (potential) presence of a breakdown.

Breakdowns in conversational agents impact users' satisfaction and their engagement~\citep{Jain:2018:DIS}. 
Previous work~\citep{Ashktorab:2019:CHI,Lee:2010:HRI,Clausen:2023:CUI} studied solutions to mitigate breakdowns in conversational agents, by focusing on strategies that can be applied by the conversational agent itself during the conversation to overcome breakdowns in a way that the user experience is impacted as little as possible.
These strategies include acknowledgment of failure, rephrasing, and clarifying that are common in human communication~\citep{Ashktorab:2019:CHI}.
\citet{Lee:2010:HRI} investigate the impact of different mitigation strategies, such as forewarning and apologizing, in the context of drink service by a robot. Their findings indicate that, in this particular context, each mitigation strategy may have a different influence on the users' experience with the robot.
In~\citep{Clausen:2023:CUI}, the authors examine users' feelings regarding different humorous repair strategies when performing a task with a smart speaker as the agent. The results of their user study show that the type of humor used by the repair strategy impacts users' experience in terms of satisfaction, perceived intelligence, and likeability of the agent. 
Repair strategies are also studied from a user perspective~\citep{Beneteau:2019:CHI,Avdic:2021:OzCHI,Alloatti:2024:CSL}, i.e., the user is the one attempting to overcome the agent's breakdowns. These types of studies commonly perform user studies to collect conversations that are manually analyzed to identify which repair strategies are used.
Unlike previous work on breakdown mitigation, our proposed method aims to automatically identify breakdowns and give insights on how to prevent them from happening, without the involvement of real users. Modifying the conversation agent based on these insights should reduce the use of repair strategies or even eliminate their need altogether. 

User simulation can serve different purposes such as the optimization of a dialogue policy using reinforcement learning~\citep{Tseng:2021:ACL,Scheffler:2001:NAACL,Schatzmann:2006:KER} and evaluation of conversational agents~\citep{Zhang:2020:KDD,Griol:2013:AAI,Wang:2023:arXiv}.
Different types of user simulators can be identified in the literature including model-based and data-driven~\citep{Balog:2023:arXiv}. 
User simulators for conversational recommendation are typically of the former type, commonly agenda-based~\citep{Zhang:2020:KDD,Sun:2018:SIGIR,Schatzmann:2007:NAACL}, which allows for an explicit control over user behavior, i.e., what actions to perform with regards to a specific goal.
Nevertheless, data-driven approaches~\citep{Lin:2021:SIGDIAL,Kreyssig:2018:SIGDIAL} have gained attention in the past years mostly due to the increase of large-scale datasets available. 
Our methodology is simulator-agnostic and can incorporate both solutions. However, due to their reliance on observed data, data-driven simulators have limitations when it comes to detecting breakdowns that happen along conversational paths that are less explored.

\begin{figure*}[t]
	\centering
	\includegraphics[width=\textwidth]{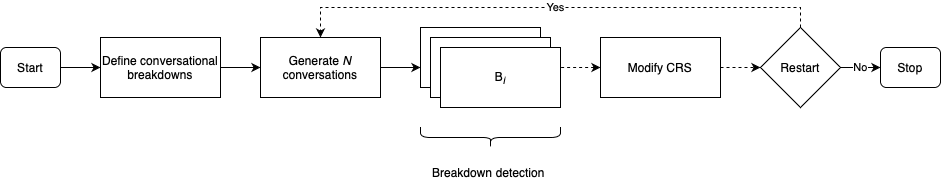}
	\caption{Flowchart of the proposed methodology, with dashed arrows denoting transitions when the methodology is used as a development tool.}
	\label{fig:flowchart}
    \Description{Flowchart displaying the four steps of the proposed methodology.}
\end{figure*}

\citet{Moller:2006:INTERSPEECH} proposed the MeMo system to automatically assess the usability of spoken dialogue systems. The system uses mental models to simulate users which can generate different types of errors. The generation of these errors is facilitated by the annotation of errors in historical dialogues. The aim of MeMo is to assess the impact of errors on the user experience rather than their detection. Indeed, the detection of errors is facilitated by the fact that they are generated by the user simulator.
In early work, \citet{Eckert:1997:ASRU} used user simulation for both automatic debugging and evaluation of conversational agents. In particular, they assess spoken dialogue systems and demonstrate that shortcomings and bugs can be identified by analyzing the length of generated dialogues.
Differing from~\citep{Eckert:1997:ASRU}, our proposed methodology is more generic (can detect multiple types of breakdowns that can be captured heuristically) and provides a more refined analysis of the breakdowns (conversational patterns based on dialogue intents).
To the best of our knowledge, no other similar work on the debugging of conversational agents has been proposed more recently.

\section{Detecting Conversational Breakdowns Using User Simulation}

This section presents the proposed methodology for the detection of conversational breakdowns along with a description and operationalization of the breakdowns considered.

\subsection{Problem Statement}
\label{sec:problem_statement}

The problem studied in this work considers a conversational recommender system (CRS), also referred to as \emph{agent}, and a user simulator (US), also referred to simply as \emph{user}, engaged in a conversation with the aim of finding items that match the user's preferences. 
Both the CRS and US are active \emph{dialogue participants} in the recommendation process.
We assume that the CRS aims to provide personalized recommendations by processing each incoming utterance and replying to it. Similarly, the US is expected to communicate the user's preferences toward specific properties or items in the form of declarations or replies to the CRS.
In our methodology, both the CRS and US are treated as ready-to-use components, i.e., no assumptions are made regarding their underlying architectures (e.g., component-based vs. end-to-end).
For our diagnostic purposes, a conversation between the CRS and the US is represented as a sequence of turns, each having an utterance accompanied by  a corresponding \emph{dialogue act}. A dialogue act comprises an intent and annotations taking the form of slot-value pairs (e.g., \texttt{REVEAL(genre=action)} is the dialogue act corresponding to the utterance ``I like action movies'').

The objective of the proposed methodology is to identify conversational breakdowns in a CRS, that is, when the flow of a conversation is discontinued or stopped.
Our methodology is designed to be generic, therefore, both the CRS and US are treated as ready-to-use components.
The detection of conversational breakdowns is performed by a set of pre-defined breakdown detection components. These components indicate whether or not their associated breakdown occurs in a given conversation.
Thus, by processing a sample of conversations between the CRS and US, they can identify recurring conversational patterns causing breakdowns.
Based on the patterns identified as problematic, improvements may be made to the CRS to make it more robust and reliable.

\subsection{Methodology}
\label{sec:method}

The proposed methodology considers three main components: the conversational recommender system, the user simulator, and breakdown detection.
Our method is divided into four steps with the last one being optional, depending on whether it is used as a diagnostic or a development tool; see Fig.~\ref{fig:flowchart}: 
\begin{enumerate}[start=0,leftmargin=0.5cm]
    \item Define conversational breakdowns to detect and implement the associated detection components. In this work, we consider three particular conversational breakdowns that are described in Section~\ref{sec:breakdown}.
    \item Generate a set of $N$ conversations between the CRS and US. The choice of $N$ may depend on the use case or the complexity of the CRS evaluated. For example, interactions with a complex CRS can exhibit more conversational paths compared to a simpler CRS, hence, $N$ should be high to explore as many paths as possible. 
	\item Run breakdown detection on the generated conversations. Each breakdown detection component $B_i$ produces a summary that includes the number of breakdowns detected per conversational pattern.
	\item (If used as a development tool) Modify the CRS based on insights from the previous step. 
    Optionally, the system developer can repeat from step (1) depending on the number of breakdowns remaining.
    In this step, it is recommended to address a single breakdown at a time (as opposed to implementing multiple changes simultaneously), as it allows for a more accurate analysis of the effect of a modification on the CRS during the next iteration. %
\end{enumerate}
This method can serve as a diagnostic tool when the CRS is a black box that cannot be modified, i.e., step (3) cannot be applied. Alternatively, it might serve as a development tool when the dialogue policy, more generally the CRS, can be modified (e.g., by updating the source code).
Indeed, by iteratively executing steps (1)--(3), one can investigate the impact of the modifications on the robustness and effectiveness of the CRS. 
We note here that step (1) could be ignored if enough historical data (i.e., recorded conversations) are available. 
Nonetheless, we advocate for the use of user simulation, as it provides the opportunity to quickly, inexpensively, and thoroughly assess the impact of the modifications made in step (3) through the generation of a large number of conversations, exploring diverse conversational paths.

\subsection{Conversational Breakdown Detection}
\label{sec:breakdown}

In this section, we describe the three specific conversational breakdowns considered in our use case. We note here that this set is by no means exhaustive and that our methodology is generic and extensible with other type of breakdowns that may be added in the future. 

\begin{figure}[t]
	\centering
	\includegraphics[width=\columnwidth]{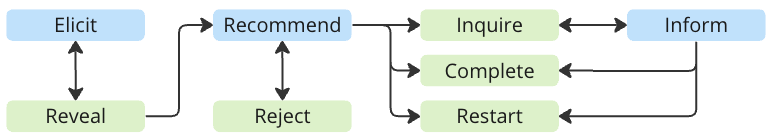}
	\caption{Simplified interaction model for conversational recommendation (inspired by \citep{Habib:2020:CIKM}). The blue and green states represent the agent and user, respectively.}
    \Description{Interaction model with 3 agent and 4 user states with arrows indicating possible transitions between the states.}
	\label{fig:dialogue_flow}
\end{figure}

\begin{itemize}
    \item[$B_1$] \emph{\textbf{System failure.}} This detector simply verifies that no system errors are thrown during conversation generation between the agent and the user. This failure can help to identify underlying bugs that were not detected during integration tests.
    
\begin{figure}[t]
    \centering
    \includegraphics[width=\columnwidth]{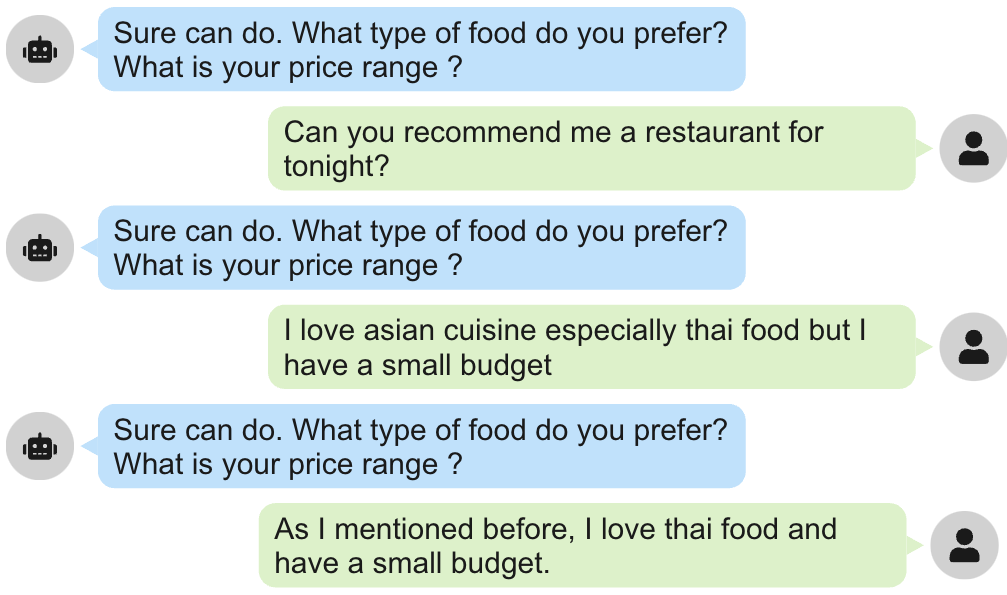}
    \caption{Example of a conversation section illustrating a \emph{dialogue of the deaf breakdown}. The green and blue bubbles represent user and agent utterances, respectively.}
    \Description{Example of conversation with 6 utterances between a user and an agent. The agent utterances are the same.}
    \label{fig:example_dod}
\end{figure}

    \item[$B_2$] \emph{\textbf{Dialogue of the deaf.}}  It identifies the agent's utterances and dialogue acts that are identical and consecutive. This failure can indicate a pitfall in the dialogue policy from which the conversational agent cannot escape. Indeed, the goal of recommending an item to the user cannot be achieved if the agent keeps repeating the same utterance for multiple turns, i.e., the conversation does not progress. Moreover, such situations can easily frustrate users, who will most likely stop interacting with the agent. An example of this breakdown is presented in Fig.~\ref{fig:example_dod}, where the agent fails to fill the slots ``type of food'' and ``price range'' with ``thai'' and ``small budget,'' respectively.
	This breakdown can be linked to the context-level error labeled \emph{repetition} in~\citep{Higashinaka:2021:SIGDIAL}, however, the definition of this error only mentions the textual similarity of utterances. The idea of repetition is also mentioned in~\citep{Moller:2007:INTERSPEECH,Oulasvirta:2006:PQS} but as a consequence of breakdown.

    \item[$B_3$] \emph{\textbf{Conversation flow discontinuation.}} This failure corresponds to an unexpected reply from either of the dialogue participants, more specifically a reply that affects negatively the naturalness of the conversation. It includes delayed replies, i.e., the participant replies to an utterance that was received at least two utterances before, and unexpected responses, i.e., a sequence of utterances with dialogue acts that should not follow each other.
    For example, considering the simplified interaction model in Fig.~\ref{fig:dialogue_flow}, which defines the flow of recommendation in terms of user/system intents, it would be unexpected behavior from the CRS to provide information about an item that the user has previously rejected as a recommendation.
	This breakdown may be classified as a task-level error by the taxonomies proposed in~\citep{Moller:2007:INTERSPEECH,Oulasvirta:2006:PQS}, as it indicates that an intent is invalid at a given state of the conversation. Additionally, we can consider the \emph{topic transition} error described in~\citep{Higashinaka:2021:SIGDIAL} as a specific case of this breakdown because it includes delayed replies.
	
\end{itemize}
Some breakdowns might be detected on the system level (\emph{system failure}), while others require semantic annotations on the utterance level (\emph{dialogue of the deaf} and \emph{conversational flow discontinuation}).
Further, we categorize \emph{system failure} as a functional breakdown, whereas \emph{dialogue of the deaf} and \emph{conversational flow discontinuation} are contextual breakdowns (cf. Table~\ref{tab:breakdowns_categories}).

\subsection{Operationalization}

In this section, we present the logic behind the implementation of detectors for the selected conversational breakdowns.

\paragraph{\textbf{System failure}}
This detector simply checks the conversation's error logs. In case an error different than \texttt{RecursionError}\footnote{The naming is Python specific, it is also known as Stack Overflow error in other languagues such as Java and C++.} is retrieved, the conversational path is flagged as problematic. \texttt{Recursi-} \texttt{onError} is associated with the \emph{dialogue of the deaf} breakdown, that is, despite a large number of interaction the participants do not understand each other to accomplish the task.

\paragraph{\textbf{Dialogue of the deaf}}
Similar to the previous detector, this one begins by checking if the conversation ended due to a \texttt{RecursionE-} \texttt{rror}, in which case, the conversational path is marked as problematic. In the absence of a \texttt{RecursionError}, it examines the agent's utterances in sequence of length three to verify if their intents are identical and texts are near identical,  indicating a breakdown. Note that using a slicing window of three utterances is an arbitrary choice and could be modified. This choice is motivated by the idea that the agent may repeat the same question twice in case of a misunderstanding, but should not persist in doing so. The pseudo-code for the detection of this breakdown is presented in Algorithm~\ref{alg:dod}.

\begin{algorithm}[t]
	\caption{Detect \emph{dialogue of deaf} breakdown}
	\label{alg:dod}
	\begin{algorithmic}[]
		\Procedure{detect\_failure}{$dialogue: Dialogue$}
		\STATE $path \gets []$
		\IF{$dialogue$ ended due to a \texttt{RecursionError}}
		\FOR{each $utterance$ in $dialogue$}
		\STATE $path$.append($utterance.intent$)
		\ENDFOR
		\STATE Return $path$
		\ENDIF
		\STATE $agent\_utterances \gets$ all agent utterances in $dialogue$
		\FOR{each $agent\_utterance$ in $agent\_utterances[:-2]$}
		\STATE $b\_same\_intent \gets$ True if $agent\_utterance.intent$ is the same intent as the one of next two agent utterances
		\STATE $b\_similar\_text \gets$ True if $agent\_utterance$ is highly similar to the next two agent utterances 
		\IF{$b\_same\_intent$ and $b\_similar\_text$}
		\FOR{each $utterance$ in $dialogue$ until $agent\_utterance$}
		\STATE $path$.append($utterance.intent$)
		\ENDFOR
		\STATE Return $path$
		\ENDIF
		\ENDFOR
		\STATE If no failure is detected, return an empty path
		\EndProcedure
	\end{algorithmic}
\end{algorithm}

\paragraph{\textbf{Conversation flow discontinuation}}
This detector is initialized with an interaction model that can be represented as a directed graph, where nodes correspond to intents and edges to allowed transitions between two intents; see the example in Fig.~\ref{fig:dialogue_flow}.
Then, following the conversation, the detector verifies if the conversational path up to the current utterance exists in this graph. For example, if the conversation has five utterances, the detector checks if the first two intents are connected, in which case, it continues by adding the third intent to the path and so on, until the last utterance. The pseudo-code in Algorithm~\ref{alg:flow_discontinuation} illustrates the logic behind this detector.

\begin{algorithm}[t]
	\caption{Detect \emph{flow discontinuation } breakdown}
	\label{alg:flow_discontinuation}
	\begin{algorithmic}[]
		\Procedure{detect\_failure}{$dialogue: Dialogue$}
		\STATE $path \gets []$
		\FOR{$utterance$ in $dialogue$}
		\STATE $path$.append($utterance.intent$)
		\IF{$path$ is not a valid path in the dialogue flow graph}
		\STATE Return $path$
		\ENDIF
		\ENDFOR
		\STATE If no discontinuity is detected, return an empty path
		\EndProcedure
	\end{algorithmic}
\end{algorithm}

\section{Case Study}
\label{sec:expsetup}

To demonstrate the feasibility and value of the proposed methodology, a case study is conducted with an existing open-sourced CRS and user simulator.
As the CRS is modifiable, we employ our methodology as a development tool to assess whether the modifications applied can decrease conversational breakdowns.

\begin{figure*}[t]
    \centering
    \begin{subfigure}{0.45\textwidth}
        \includegraphics[width=\textwidth]{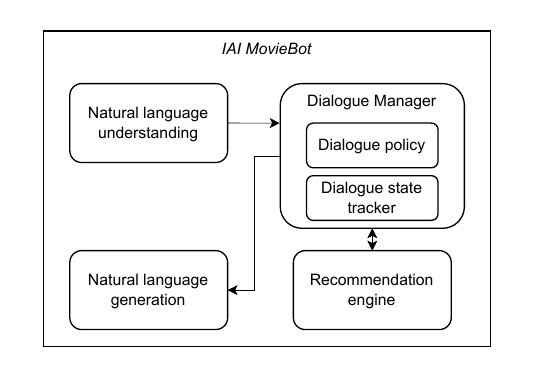}
        \caption{\emph{IAI MovieBot}}
        \label{fig:architecture:crs}
    \end{subfigure}
    \hfill
    \begin{subfigure}{0.45\textwidth}
        \includegraphics[width=\textwidth]{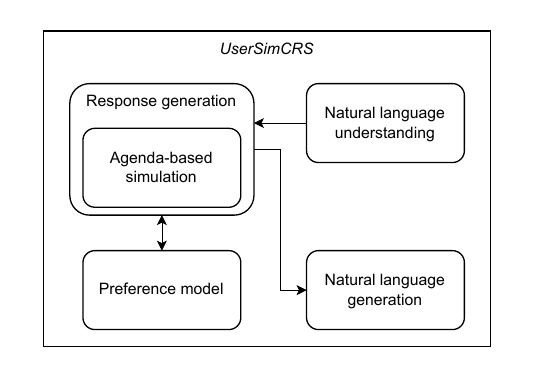}
        \caption{\emph{UserSimCRS}}
        \label{fig:architecture:usersim}
    \end{subfigure}
    \caption{Architecture of dialogue participants: \emph{IAI MovieBot} and \emph{UserSimCRS}.}
    \Description{The subfigure labelled (a) and (b) present the architecture of IAI MoveiBot and UserSimCRS. Each comprises 4 components and arrows between them represents the circulation of an utterance upon reception. The components of IAI MoveiBot are: natural language understanding, dialogue manager, recommendation engine, and natural language generation. The components of UserSimCRS are: natural language understanding, response generation, preference model, and natural language generation.}
    \label{fig:architecture}
\end{figure*}

\emph{IAI MovieBot}~\citeanon{Habib:2020:CIKM} is a conversational recommender system in the movie domain.
It has a component-based architecture with four core components (Fig.~\ref{fig:architecture:crs}), namely, natural language understanding (NLU), dialogue manager (comprising dialogue policy and dialogue state tracking), recommendation engine, and natural language generation, that is very similar to the typical architecture found in~\citep{Jannach:2021:CSUR}. 
The user utterance is processed by a rule-based NLU component. Then, the structured representation of the utterance is received by the dialogue manager that is responsible for deciding the next action of \emph{IAI MovieBot} based on a pre-defined set of rules. It is connected to the recommendation engine, which is responsible for finding relevant items with regards to the dialogue state. 
Finally, the natural language generation component sends a template-based response to the user based on the next action chosen by the dialogue manager.
For user simulation, we use \emph{UserSimCRS}~\citeanon{Afzali:2023:WSDM} that is designed for evaluation of the CRSs and implements an agenda-based simulator. Its architecture is based on a typical task-oriented dialogue system (Fig.~\ref{fig:architecture:usersim}), and thus is relatively similar to that of \emph{IAI MovieBot}, the main difference is the absence of recommendation engine.
Two practical differences with \emph{IAI MovieBot} can be pointed out. First, the NLU component is trained on a small corpora of dialogues between users and \emph{IAI MovieBot}. Second, although the NLG component also uses templates, they are extracted from the corpora of dialogues, as opposed to being pre-defined.
Agenda-based simulation assumes a Markovian state representation; additionally, the allowed state transitions are specified by an interaction model that is designed specifically for the task of conversational recommendations.
The simulator determines the next action based on the agenda and the current context of the conversation.

\begin{figure*}[t]
	\centering
	\includegraphics[width=\textwidth]{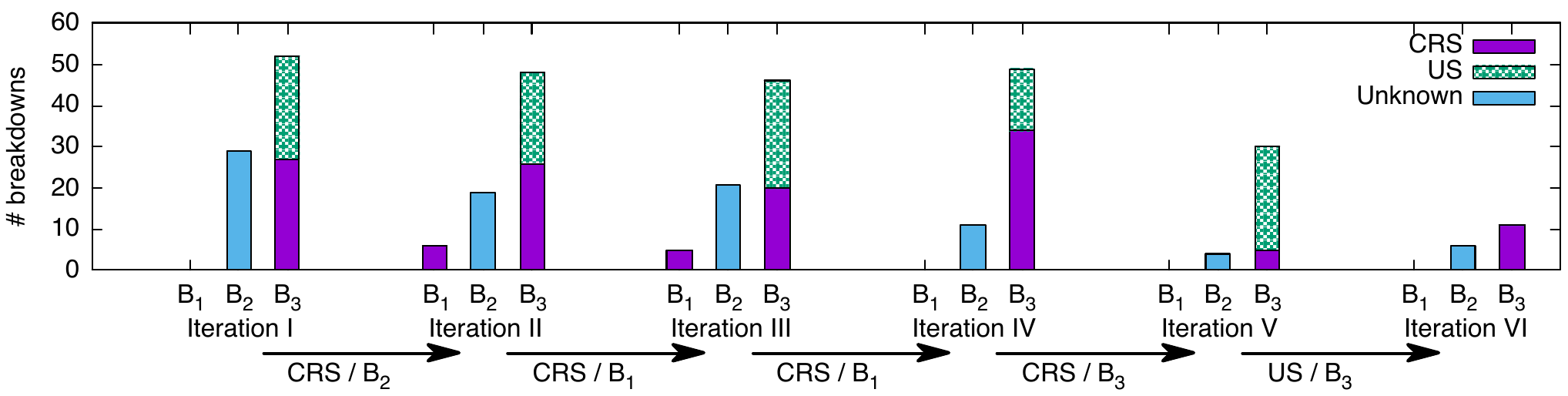}
	\caption{Conversational breakdowns per type for each iteration (groups). $B_1$, $B_2$, and $B_3$ represent system failure, dialogue of the deaf, and flow discontinuation, respectively. Colors indicate whether the breakdowns can be attributed to the CRS (purple), the US (green), or cannot be attributed (blue). Our aim in this work is to decrease the number of breakdowns especially those attributed to the CRS. The arrows between the iterations indicate which participant (CRS/US) was modified and what breakdown was targeted.}
    \Description{Bar chart of number of breakdowns per iteration and per breakdown. A description of the variation in number of breakdown per iteration is described in Section 4.}	
 \label{fig:results}
\end{figure*}

For this case study, we generate $N=100$ conversations per iteration\footnote{The generated conversations are made publicly available at: \url{https://github.com/NoB0/crs-breakdown-detection}} as the action space of \emph{IAI MovieBot} is relatively simple (i.e., 10 possible actions~\citeanon{Habib:2020:CIKM}). %
We perform five iterations; in each, we update \emph{IAI MovieBot} by addressing one specific type of breakdown. Additionally, we conduct an extra iteration where we use an updated version of \emph{UserSimCRS}.
Fig.~\ref{fig:results} presents a summary of the different types of breakdowns detected in each iteration. Moreover, the figure provides an overview of the breakdowns addressed in each iteration.
We expect the number of breakdowns related to the one in focus to decrease. However, it is possible that the modifications applied lead to an increase or decrease in the other types of breakdowns.
Additionally, we attempt to attribute each breakdown to either the CRS or the US as indicated by color-coding in Fig.~\ref{fig:results}. 
By analyzing the logs, we can find the source of a $B_1$ breakdown, while for $B_3$, we investigate the conversational path, with regards to intents, leading to a breakdown to identify which participant made a ``forbidden'' transition, i.e., one that is not allowed by the interaction model. For example, the US accepts a recommendation (i.e., intent \texttt{ACCEPT}) while the CRS is eliciting the US's preferences (i.e., intent \texttt{ELICIT}). The responsibility for $B_2$ cannot be attributed to a specific participant as both participants fail to escape the misunderstanding loop they are in despite rephrasing utterances.
We note here that the \emph{UserSimCRS} is imperfect as illustrated in the responsibility distribution of $B_3$ breakdowns.
\begin{enumerate}[label=\textnormal{(\Roman*)}]
	\item First, we perform conversational breakdown detection with the original version of \emph{IAI MovieBot}. For this iteration, we notice that both participants (i.e., \emph{IAI MovieBot} and \emph{UserSimCRS}) are free of system failure, while a significant number of $B_2$ and $B_3$ breakdowns are detected. Therefore, we first decide to focus on reducing the number $B_2$ breakdowns by modifying \emph{IAI MovieBot}.
	During the analysis of the simulated conversations, we note that this breakdown often appears when \emph{IAI MovieBot} does not understand \emph{UserSimCRS} and does not manage to find a way to pick up the conversation, as illustrated in Fig.~\ref{fig:example_dod_study}. Hence, we apply a straightforward modification that consists in \emph{IAI MovieBot} proactively restarting the conversation, using a predefined message, when such situation happens.
    \begin{figure}[t]
        \centering
        \includegraphics[width=\columnwidth]{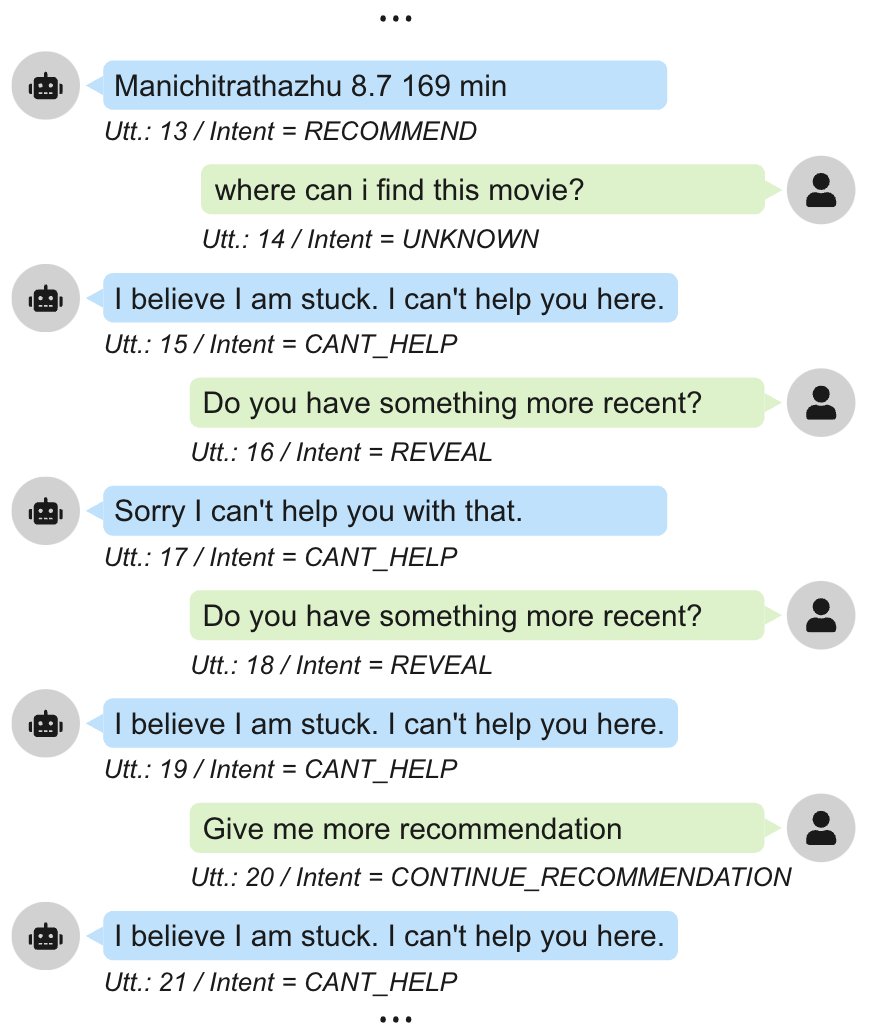}
        \caption{Excerpt of a conversation between \emph{UserSimCRS} (in green) and \emph{IAI MovieBot} (in blue) illustrating a $B_2$ breakdown, i.e., \emph{IAI MovieBot} does not understand that \emph{UserSimCRS} asks for an additional recommendation. In the entire conversation, we count the \texttt{CANT\_HELP} intent 12 times in a row.}
        \Description{Excerpt of a conversation with five utterances from IAI MovieBot and four from UserSimCRS. The number of the utterance and its intent is display below its bubble text. Four out of the five utterances from IAI MovieBot have the intent CANT_HELP.}
        \label{fig:example_dod_study}
    \end{figure}

	\item As expected, during the second iteration, we observe a decrease in the number of $B_2$ breakdowns. However, we note that some $B_1$ breakdowns are detected in \emph{IAI MovieBot}. One possible explanation is that new conversational paths can be explored as \emph{IAI MovieBot} is more robust to $B_2$ breakdowns. After looking at the summary of the breakdowns and the logs, we decide to fix a bug that appears when \emph{IAI MovieBot} tries to conclude the conversation.

	\item In this iteration, we observe a decrease in the number of $B_1$ breakdowns caused by \emph{IAI MovieBot}. Interestingly, we find that the number of $B_3$ breakdown also decreases, while $B_2$ breakdowns slightly increase. We continue to focus on $B_1$ breakdown, hence, we fix a new bug discovered when \emph{IAI MovieBot} wants to remove a preference from the user model.
	
	\item During this iteration, we do not detect $B_1$ caused by \emph{IAI MovieBot}. This shows that the previous modifications helped to improve its the robustness. Unlike the previous iteration, the number of $B_3$ breakdowns increases, while we detect less $B_2$ breakdowns. 
	For the last modification of \emph{IAI MovieBot}, we shift our focus towards $B_3$ breakdowns. The analysis of the simulated conversations from the previous iterations reveals that this type of breakdown can appear when \emph{IAI MovieBot} fails to appropriately reply to \emph{UserSimCRS}'s utterance accepting a recommendation that is different from the predefined message ``I like this recommendation.'' Some examples of such utterances are: ``Sounds good I like it,'' ``Sounds good'' and ``Ok I like this recommendation.'' Fig.~\ref{fig:flow_discontinuation_error} shows partial conversation flows after a recommendation is made; three out of the five flows observed are not allowed in the interaction model. 
	Our approach to reduce $B_3$ breakdowns in the aforementioned situation is to improve the recognition of the \texttt{ACCEPT} intent by the natural language understanding component of \emph{IAI MovieBot}.
    \begin{figure}[t]
    	\centering
    	\includegraphics[width=\columnwidth]{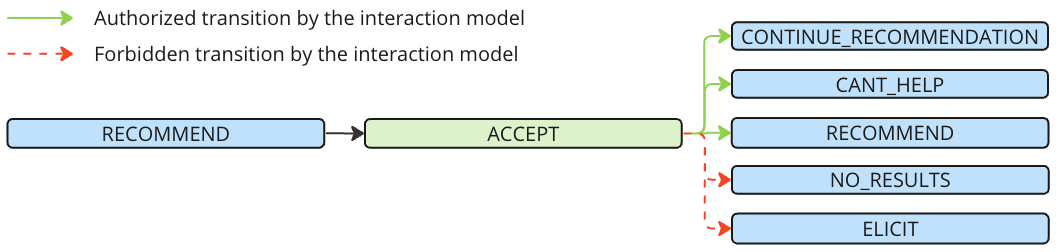}
    	\caption{Partial conversational flows observed in simulated conversations after a recommendation is made by \emph{IAI MovieBot}. The intent of utterances from \emph{IAI MovieBot} and \emph{UserSimCRS} are represented in blue and green, respectively.}
        \Description{Flowchart representing the agent states observed after the transition RECOMMEND-ACCEPT. The transitions marked with red dash line are forbidden by the interaction model, while the ones in green lines are authorised.}
    	\label{fig:flow_discontinuation_error}
    \end{figure}
    
	\item The addition of new rules, i.e., patterns such as ``like it,'' ``love it,'' and ``i like this,''  to improve the recognition of the \texttt{ACCEPT} intent results in a major drop of $B_3$ breakdowns caused by  \emph{IAI MovieBot}. However, the number of breakdowns caused by \emph{UserSimCRS} increases. Therefore, we update \emph{UserSimCRS} with a focus on $B_3$ breakdowns, more especially on \emph{UserSimulation}'s agenda update. 
	Originally, an agenda is initialized for the entire conversation and the next action is pulled off when the CRS replies in an expected manner, otherwise it is sampled~\citeanon{Afzali:2023:WSDM}. However, we find that this could lead to $B_3$, even if \emph{IAI MovieBot} responds appropriately. For example, the intent \texttt{REVEAL} may be followed by \texttt{ACCEPT} in the agenda; yet, it may happen that no results were found after \texttt{REVEAL}, in which case it would not make sense to have \texttt{ACCEPT} as the user response. Therefore, we update \emph{UserSimCRS} such that the the next action is always sampled.
	
	\item  In this extra iteration, we observe a minor increase of $B_2$ and $B_3$ breakdowns. Interestingly, $B_3$ breakdowns caused by \emph{UserSimCRS} are eliminated entirely after our last update. 
\end{enumerate}
\begin{table*}[t]
    \centering
    \caption{Analysis of conversational paths explored during each iteration. A conversation is deemed successful if at least one recommendation made by the CRS is accepted. Existing and new conversational paths are identified by comparison of conversational paths from the previous and current iterations. The average length of the conversations is calculated based on the number of utterances.}
    \label{tab:path_stats}
    \begin{tabular*}{\textwidth}{@{\extracolsep{\fill}}lcccccc}
    \hline
         & \textbf{\# Unique conv. path} & \textbf{Avg. conv. length} & \multicolumn{2}{c}{\textbf{Existing conv. path}} &  \multicolumn{2}{c}{\textbf{New conv. path}}\\ 
         \cline{4-5} \cline{6-7}
         & & & Success & Not Success & Success & Not Success \\ \hline
         Iteration \rom{1} & 96 & 19.84 $\pm$ 14.45 & & & 66 & 34  \\
         Iteration \rom{2} & 92 & 19.34 $\pm$ 11.02 & 7 & 2 & 59 & 32 \\
         Iteration \rom{3} & 93 & 19.29 $\pm$ 10.87 & 5 & 7 & 55 & 33 \\
         Iteration \rom{4} & 98 & 19.42 $\pm$ 12.97 & 5 & 3 & 69 & 23 \\
         Iteration \rom{5} & 78 & 17.6 $\pm$ 10.77 & 17 & 7 & 56 & 20 \\
         Iteration \rom{6} & 76 & 15.04 $\pm$ 9.02 & 26 & 4 & 31 & 39 \\ \hline
    \end{tabular*}
\end{table*}

\noindent
Table~\ref{tab:path_stats} provides an analysis of the conversational paths explored across different iterations.
We notice that in iterations I--IV there is a large number of number of unique conversational paths explored, which reduces later in iterations V and VI. 
Moreover, we see that the length of the conversations follows a similar pattern, i.e., it decreases after iteration IV. While there is no direct explanation for this, one hypothesis is that the modifications applied lead to more direct conversations, that is, less back and forth between the participants to give feedback for a recommendation. Additionally, the high standard deviation of conversation length hints at the diversity of the conversational paths explored.
When examining the success of these conversational paths, i.e., both existing and new paths, we make three main observations.
First, there is a certain overlap between the conversational paths explored from one iteration to the next, but the majority of paths explored are new.
Second, in most cases, more successful conversational paths are kept from one iteration to the next. Indeed, there is a higher proportion of successful paths within the existing ones. 
Third, we see that the number of new and unsuccessful paths decreases up until the last iteration. It indicates that the modifications applied to \emph{IAI MovieBot} improve its ability to satisfy \emph{UserSimCRS}'s needs. However, this is not the case after the update of \emph{UserSimCRS} in the last iteration.
The aforementioned observations support the idea that \emph{IAI MovieBot} is becoming more robust and better at satisfying \emph{UserSimCRS}'s needs with each iteration. Indeed, the modifications applied to \emph{IAI MovieBot} lead to the exploration of diverse and new conversational paths, while providing better recommendations in different contexts. 

In summary, this case study demonstrates that our methodology can be used to trace down different types of conversational breakdowns and to make \emph{IAI MovieBot} more robust to them in a few iterations.
Indeed, in two iterations (iterations \rom{2} and \rom{3}), we eliminate the \emph{system failures} caused by \emph{IAI MovieBot}. Furthermore, the modifications applied during iterations \rom{1} and \rom{4} illustrate that \emph{IAI MovieBot} can be improved in a more straightforward manner thanks to the identification of problematic conversational flows.
It is noteworthy that each modification might also affect other types of breakdowns. For example, the modification applied to tackle \emph{system failures} in iteration \rom{3} increases the number of \emph{flow discontinuation} detected in iteration \rom{4}. This can be explained by the fact that after a fix other conversational paths might be explored and new breakdowns can be uncovered.
The analysis of the conversational patterns leading to \emph{flow discontinuation} breakdowns shows that \emph{UserSimCRS} is responsible for a large number of breakdowns. This highlights that despite being used to evaluate one CRS, the proposed methodology may also benefit the user simulator in parallel to improve its robustness and expose its limitations. Indeed, the last iteration shows that improving the quality of \emph{UserSimCRS} affects the results.
We note here that \emph{flow discontinuation} breakdowns caused by \emph{IAI MovieBot} are still detected, mostly in the situation where \emph{UserSimCRS} acknowledges after an utterance from \emph{IAI MovieBot}.

\section{Discussion}
\label{sec:disc}

One main concern related to the use of user simulation for the evaluation of CRSs regards its quality; that is, the user simulator is most likely imperfect (e.g., overly simplified user model or context). Indeed, in the case of evaluation, the quality of the user simulation can impact the interpretation of the results obtained as demonstrated in the case study. 
Assuming that the user simulator can be controlled or modified (e.g., by updating its source code), it can be developed in parallel with the CRS to reduce the number of breakdowns and enhance its overall quality as previously argued in~\citeanon{Afzali:2023:WSDM}. 
Having a more robust user simulator can lead to the exploration of new conversational paths, therefore, ensuring a better coverage of breakdown-free interaction space.
We also note that our method would greatly benefit from future advances related to the open challenge of fairly representing all types of users in simulation~\citep{Balog:2023:arXiv}. Indeed, the analysis of conversational breakdown detection would be more comprehensive by simulating users outside the target user population as well. 

A discerning reader may also ask the following question: if the dialogue policy is trained using reinforcement learning (RL), which relies on user simulation, what benefits does simulation offer for diagnosis? 
Even if the very same user simulator is employed in both cases, it can still serve as a diagnostic tool, ensuring that the dialogue policy has been trained sufficiently to handle various conversational paths effectively.
Nonetheless, there is no requirement to use an identical user simulator.
RL often employs data-driven user simulators, whereas we argue that a model-based simulator is more suitable for our objectives as it provides greater control over user behavior.

The conversational breakdown detectors described in this work are rather generic and are meant to provide a starting point to illustrate and validate our methodology. 
Indeed, these detectors can be used to test any type of CRS agent (i.e., component-based and end-to-end) as long as semantic annotations are available for user and agent utterances.
One could also design new detectors that target more refined breakdowns, do not need semantic annotations, or are more specific to recommendation goals (e.g., recommending a single item vs. a set of items).
A possible approach to develop these new detectors is to exploit previous observations/feedback from experts, beta testers, or users (if the CRS is deployed). Then, one can devise heuristics to capture the identified issues automatically. For example, if a CRS works with slot-value pairs to represent preferences and the natural language understanding component shows some imperfections, a naive conversational breakdown could be the CRS asking preferences for the same slot over and over.

A major advantage of our methodology is that it is generic and can be used with any CRS architecture. 
There are, however, differences across CRSs in how the detected breakdowns may be addressed. In case of component-based systems, the dialogue policy (or its equivalent that is responsible for the conversational behavior of the system) needs to be changed directly; in a rule-based system this entails updating the source code, while in an RL-based system with a learned policy, it requires a re-training of the policy with an updated user simulator.  As for end-to-end systems, where the system's behavior cannot be controlled directly but only via the training examples it is exposed to, addressing breakdowns might require the collection of new training examples (possibly with the help of user simulation) with the desired system behavior.

\section{Conclusion}
\label{sec:concl}

In this work, we have proposed a methodology that leverages user simulation to identify breakdowns in conversational recommender systems.
It consists of the definition of conversational breakdowns and the implementation of their associated detectors. These detectors extract  conversational paths (i.e., sequence of dialogue intents) leading to the related breakdowns.
The detection of breakdowns is performed on a sample of conversations between the conversational recommender system (CRS) and user simulator (US). We capitalize on the fact that user simulation is a fast and cheap way of collecting a large sample of conversations to be analyzed.
The methodology can be used as a diagnostic or a development tool. In the latter case, one can iteratively improve the CRS to reduce the number of breakdowns, thereby making it more robust. 
We have demonstrated this in a case study with an existing CRS and US. Indeed, in five iterations, we have successfully removed all \emph{system failures} caused by the CRS and drastically reduced the number of other types of breakdowns.
The case study also highlights the fact that the US is imperfect and may be responsible for some portion of the conversational breakdowns detected. In an extra iteration, we have shown that the methodology can also be used to improve the robustness of the US in parallel with the CRS.

Future directions include the application of our method to other CRSs and USs to ensure its adaptability.  
Furthermore, new conversational breakdown detectors can be created to provide a more extensive test of the CRS.  This could include, for example, the detection of off-topic utterances that can disrupt the flow of the conversation.

\begin{acks}
This research was supported by the Norwegian Research Center for AI Innovation, NorwAI (Research Council of Norway, project number 309834).
\end{acks}

\bibliographystyle{ACM-Reference-Format}
\bibliography{cui2024-crs-breakdowns.bib}

\end{document}